\documentclass[conference]{IEEEtran}
\usepackage[cmex10]{amsmath}

\ifCLASSINFOpdf
  \usepackage[pdftex]{graphicx}
\else
\usepackage[dvips]{graphicx}
\fi
\begin{document}

\title{Multi-Edge type Unequal Error Protection LDPC Codes}

\author{\IEEEauthorblockN{H. V. Beltr\~{a}o Neto, W. Henkel}
\IEEEauthorblockA{Jacobs University Bremen\\
Campus Ring 1\\D-28759 Bremen, Germany\\
Email: \{h.beltrao, w.henkel\}@jacobs-university.de\\} \and
\IEEEauthorblockN{V. C. da Rocha Jr.} \IEEEauthorblockA{Department
of Electronics and Systems\\ Federal University of
Pernambuco\\ P.O. Box 7800 Recife, Brazil, 50711-970\\
Email: vcr@ufpe.br}\\ }

\maketitle

\begin{abstract}
Irregular low-density parity check (LDPC) codes are particularly
well-suited for transmission schemes that require unequal error
protection (UEP) of the transmitted data due to the different
connection degrees of its variable nodes. However, this UEP
capability is strongly dependent on the connection profile among the
protection classes. This paper applies a multi-edge type analysis of
LDPC codes for optimizing such connection profile according to the
performance requirements of each protection class. This allows the
construction of UEP-LDPC codes where the difference between the
performance of the protection classes can be adjusted and with an
UEP capability that does not vanish as the number of decoding
iterations grows.
\end{abstract}

\section{Introduction} 
In communication systems where source bits with different
sensitivities to errors are being transmitted, it is often wasteful
or even infeasible to provide uniform protection for the whole data
stream. In this scenario, the common strategy is the use of schemes
with unequal error protection (UEP) capabilities. There are mainly
three strategies to achieve UEP on transmission systems: bit
loading, multilevel coded modulation, and channel coding
\cite{Henkel:hindawi}. In this paper, we will focus on the latter,
more specifically on low-density parity-check (LDPC) codes that
provide inherent unequal error protection within a codeword.

Irregular LDPC codes \cite{Richarson:irregLDPC} can inherently
provide unequal error protection due to the different connection
degrees of the coded bits. The connection degrees of the variable
and check nodes of such codes are defined by the polynomials
$\lambda(x)=\sum_{i=2}^{d_{v_{max}}}\lambda_ix^{i-1}$ and
$\rho(x)=\sum_{i=2}^{d_{c_{max}}}\rho_ix^{i-1}$, where $d_{v_{max}}$
and $d_{c_{max}}$ are the maximum variable and check node degrees of
the code. From now on, we will refer to irregular LDPC codes where
the variable nodes are divided into disjoint sets called protection
classes as unequal error protection LDPC codes (UEP-LDPC).

A flexible optimization of the irregularity profile of irregular
LDPC codes based on a hierarchical optimization of the variable node
degree distribution was proposed in \cite{Poulliat:UEPLDPC}, where
the authors interpret the UEP properties of an LDPC code as
different local convergence speeds, i.e., the most protected bits
are assigned to the bits in the codeword which converge to their
right value in the smallest number of iterations. This assumption is
made in order to cope with the observation that the UEP gradation
vanishes as the number of iterations grow, a fact also observed in
\cite{plotkin:UEP}. In \cite{ldpcuep}, the authors observed that
this vanishing UEP gradation of an iteratively decoded LDPC code  is
dependent on the algorithm used to construct the parity check
matrix, and suggested that the connectivity between the classes is
the key factor to be observed if the UEP capabilities should be held
as the number of iterations grows.
\newline \indent Herein, we propose an optimization algorithm for the connectivity profile between the
different protection classes of LDPC codes in order to not only keep
the UEP capability of a code for a moderate to large number of
decoding iterations, but also to adjust the performance of the
protection classes as required for different applications. This is
achieved by means of a multi-edge type (MET) analysis
\cite{RiU04b/LTHC,mct} of the LDPC codes. The multi-edge analysis
enables us to distinguish between the messages exchanged during the
iterative decoding among the different protection classes within one
codeword. Thus, we can control the amount of information that the
most protected classes receive from the less protected ones and vice
versa. If the most protected classes receive a lot of information
from the less protected ones, its performance will be decreased
while the one of the less protected classes will be enhanced. Our
main goal is to show how this exchange of performance among the
protection classes can be controlled and optimized.\newline \indent
This paper is organized as follows. In Section II, we describe the
multi-edge type analysis of UEP-LDPC codes. Section III discusses
the asymptotic analysis of multi-edge type UEP-LDPC codes and the
optimization algorithm used to optimize the connection profile
between the protection classes. In Section IV, we show the results
of the developed optimization method for a chosen example. Finally,
some concluding remarks are drawn in Section V.

\section{Multi-Edge type Unequal Error Protection LDPC Codes}
\subsection{Multi-edge LDPC codes}
Multi-edge type LDPC codes \cite{RiU04b/LTHC} are a generalization
of irregular and regular LDPC codes. Diverting from standard LDPC
ensembles where the graph connectivity is constrained only by the
node degrees, in the multi-edge setting, several edge classes can be
defined and every node is characterized by the number of connections
to edges of each class. Within this framework, the code ensemble can
be specified through two multinomials associated to variable and
check nodes. The two multinomials are defined by \cite{mct}
\begin{equation}
L(\textbf{r},\textbf{x})=\sum
L_{\textbf{b},\textbf{d}}\textbf{r}^\textbf{b}\textbf{x}^\textbf{d},
\end{equation}
\begin{equation}
R(\textbf{x})=\sum R_\textbf{d}\textbf{x}^\textbf{d},
\end{equation}
where \textbf{b},~\textbf{d},~\textbf{r}, and \textbf{x} are vectors
which are explained as follows. First, let $m_e$ denote the number
of edge types used to represent the graph ensemble and $m_r$ the
number of different received distributions. The number $m_r$
represents the fact that the different bits can go through different
channels and thus, have different received distributions. Each node
in the ensemble graph has associated to it a vector
$\textbf{x}=(x_1,...,x_{m_{e}})$ that indicates the different types
of edges connected to it, and a vector
$\textbf{d}=(d_1,...,d_{m_{e}})$ referred to as \textit{edge degree
vector} which denotes the number of connections of a node to edges
of type $i$, where $i\in(1,\ldots,m_e)$.

For the variable nodes, there is additionally the vector
$\textbf{r}=(r_1,...,r_{m_{r}})$ which represents the different
received distributions\footnote{In the multi-edge framework, one can
consider that the different variable node types may have different
received distributions, i.e., the associated bits may be transmitted
through different channels. In this work, we consider that the
variable nodes have access solely to one observation and that the
transmission is made through an AWGN channel.}, and the vector
$\textbf{b}=(b_0,...,b_{m_{r}})$ that indicates the number of
connections to the different received distributions ($b_0$ is used
to indicate the puncturing of a variable node). In the sequel, we
assume that $\textbf{b}$ has exactly one entry set to 1 and the rest
set to zero.  This simply indicates that each variable node has
access to only one channel observation at a time. We use
$\textbf{x}^\textbf{d}$ to denote $\prod_{i=1}^{m_e}x_i^{d_{i}}$ and
$\textbf{r}^\textbf{b}$ to denote $\prod_{i=0}^{m_r}r_i^{b_{i}}$.
Finally, the coefficients $L_{\textbf{b},\textbf{d}}$ and
$R_\textbf{d}$ are non-negative reals that represent the fraction of
variable nodes of type (\textbf{b},\textbf{d}) and check nodes of
type (\textbf{d}) within a codeword, respectively. Furthermore, we
have the additional notations defined in \cite{mct}
\begin{equation}
L_{r_i}(\textbf{r},\textbf{x})=\frac{\mathrm{d}L(\textbf{r},\textbf{x})}{\mathrm{d}
r_i},
\end{equation}
\begin{equation}
L_{x_i}(\textbf{r},\textbf{x})=\frac{\mathrm{d}L(\textbf{r},\textbf{x})}{\mathrm{d}
x_i},
\end{equation}
\begin{equation}
R_{x_i}(\textbf{x})=\frac{\mathrm{d} R(\textbf{x})}{\mathrm{d} x_i}.
\end{equation}

Note that in a valid multi-edge ensemble, the number of connections
of each edge type should be the same at both variable and check
nodes sides. This gives rise to the \textit{socket count equality}
constraint which can be written as
\begin{equation}
L_{x_i}(\textbf{1},\textbf{1})=R_{x_i}(\textbf{1}),~~~~i=1,...,m_e,
\end{equation}
where $\textbf{1}$ denotes a vector with all entries equal to 1,
with length being clear from the context.
\begin{figure}
\centering
    \includegraphics[scale=0.65]{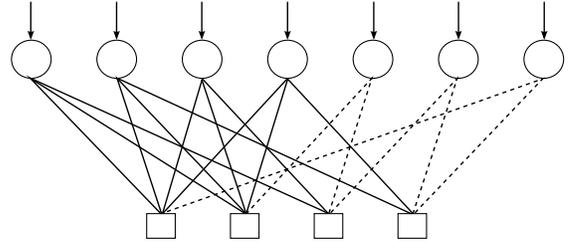}\\
  \caption{Multi-edge graph with two different edge types and one received distribution.}\label{2edgefg}
\end{figure}

Unequal error protection LDPC codes can be included in a multi-edge
framework in a straightforward way. This can be done by
distinguishing between the edges connected to different protection
classes within a codeword. According to this strategy, the edges
connected to variable nodes within a protection class are all of the
same type. For example, in Fig. \ref{2edgefg} the first 4 variable
nodes can be seen as one protection class since they are connected
only to type 1 edges (depicted by solid lines), and the last 3
variable nodes compound another protection class, since they are
only connected to type 2 edges (depicted by the dashed lines). It is
worth noting that opposed to the variable nodes, the check nodes
admit connections with edges of different types simultaneously as
can be inferred from Fig.\ref{2edgefg}. In the following, we will
divide the variable nodes into $m_e$ protection classes
$(C_1,C_2,\ldots,C_{m_e})$ with degrading levels of protection.

\subsection{Edge perspective notation}
The connection between the protection classes occurs through the
check nodes since they can have different types of edges attached to
them. Consider irregular LDPC codes with node perspective variable
and check node multi-edge multinomials
$L(\textbf{r},\textbf{x})=\sum
L_{\textbf{b},\textbf{d}}\textbf{r}^\textbf{b}\textbf{x}^\textbf{d}$
and $R(\textbf{x})=\sum R_\textbf{d}\textbf{x}^\textbf{d}$,
respectively.

In order to implement the optimization algorithm, it will be more
convenient to work with the edge, instead of the node perspective.
We now define edge perspective multi-edge multinomials. Let
$\rho_{\textbf{d}}^{(j)}$ denote the fraction of type $j$
($j=1,\ldots,m_e$) edges connected to check nodes of type
\textbf{d}. The fraction $\rho_{\textbf{d}}^{(j)}$ is calculated
with respect to the total number of edges of type $j$ and is thus
given by
\begin{equation}
\rho_{\textbf{d}}^{(j)}=\frac{d_jR_{\textbf{d}}}{R_{x_{j}}(\textbf{1})}.
\label{rho:local}
\end{equation}
Similarly, let $\lambda_i^{(j)}$ denote the fraction (computed with
respect to the total number of type $j$ edges) of type $j$ edges
connected to variable nodes of degree $i$. This gives us the
following edge perspective multinomial
\begin{equation}
\lambda^{(j)}(x)=\sum_i\lambda_i^{(j)}x^{i-1}, \label{locallambda}
\end{equation}
where $\lambda_i^{(j)}$ represents the fraction of class $j$ edges
connected to variable nodes with degree $i$. In the next section, we
will use Eqs. (\ref{rho:local}) and (\ref{locallambda}) in the
derivation of the optimization algorithm for the connection profile
among the protection classes of an UEP-LDPC code.

\section{Check node profile optimization}
\subsection{Asymptotic Analysis}
Our main objective is, given the overall variable and check node
degree distributions of an UEP-LDPC code, to optimize the connection
profiles between the different protection classes in order to
control the amount of protection of each class while preserving the
UEP capability of the code after a moderate to high number of
decoding iterations. In order to simplify the optimization
algorithm, we suppose that the LDPC code to be optimized is
check-regular, i.e., all the check nodes have the same degree.

Despite of having the same degree, each check node may have a
different number of edges belonging to each one of the $m_e$
classes. Consider for example a check node with an associated edge
degree vector $\textbf{d}=(d_1,d_2,\ldots,d_{m_e})$, where $d_i$ is
the number of connections to the protection class $i$ and
$\sum_{i=1}^{m_e}d_i=d_{c_{max}}$. If we then consider a code with
$m_e=3$ protection classes, each check node may be connected to
$d_1$ edges of class 1, $d_2$ edges of class 2, and $d_3$ edges of
class 3. This posed, one can compute the evolution of the iterative
decoding by means of density evolution.

We assume here standard belief propagation decoding of LDPC codes
where the messages exchanged between the variable and check nodes
are log-likelihood ratios having a symmetric Gaussian distribution
(variance equals twice the mean). Before dealing with the subtle
case of irregular multi-edge type LDPC codes, let's make for now two
simplifying assumptions: 1) All variable nodes have the same degree
$d_v$. 2) All check nodes have the same edge degree vector
$\textbf{d}$.

Under these assumptions, we can generalize the results obtained in
\cite{EXITfunc:reg} for regular LDPC codes and compute the mutual
information  between the output message of a class $C_j$,
degree-$d_v$ variable node and the observed channel
value\footnote{For the sake of simplicity, through the rest of text
we will simply speak of ``mutual information of some variable" when
we refer to the mutual information between this variable and its
observation through the channel.} at iteration $l$, which is given
by
\begin{equation}
I_{v,l}^{C_j}=J\left(\sqrt{4/\sigma^2+(d_v-1)[J^{-1}(I_{c,{l-1}}^{C_j})]^2}\right),
\label{Iv:reg}
\end{equation}
\noindent where $\sigma^2$ is the channel noise variance,
$I_{c,l-1}^{C_j}$ is the mutual information received from the
neighboring check nodes at iteration $l-1$ and the $J(.)$ function
is defined as in \cite{J:brink}
\begin{equation}
J(\sigma)=1-\int_{-\infty}^{\infty}\frac{e^{\frac{(\xi-\sigma^2/2)^2}{2\sigma^2}}}{\sqrt{2\pi}\sigma}\cdot\log_2[1+e^{-\xi}]d\xi~.
\label{J}
\end{equation}

\noindent Similarly, the mutual information sent to the variable
nodes of class $C_j$ from a check node with an associated vector
$\textbf{d}$ can be expressed as
\begin{align}
&I_{c,l}^{C_j}=1\nonumber\\
&-J\left(\sqrt{(d_j-1)(J^{-1}(1-I_{v,l}^{C_j}))^2+\sum_{i\ne
j}d_i(J^{-1}(I_{v,l}^{C_i}))^2 }\right). \label{Ic:reg}
\end{align}

Let's now deal with the more general case of irregular codes. For
irregular codes, the incoming densities to a node are not
necessarily equal due to varying degrees. Therefore, the overall
mutual information at the output of the variable and check nodes is
averaged over the different degrees. Also note that, in the
irregular case, for optimizing the connection profile between the
protection classes, we need to consider the case where check nodes
with different edge degree vector $\textbf{d}$ are allowed. This
posed, the mutual information between the received channel values
and  the messages sent from the check to variable nodes and from
variable to check nodes at iteration $l$ within the protection class
$C_j$, computed by means of density evolution using the Gaussian
approximation \cite{EXITfunc:irreg}, are given by
\begin{equation}
I_{v,l}^{C_j}=\sum_{i=2}^{d_{v_{max}}}\lambda_i^{(j)}J\left(\sqrt{4/\sigma^2+(i-1)[J^{-1}(I_{c,{l-1}}^{C_j})]^2}\right),
\label{Iv:irreg}
\end{equation}
\begin{align}
&I_{c,l}^{C_j}=1-\sum_{s=1}^{d_{c_{max}}}\sum_{\textbf{d}:d_j=s}\rho_{\textbf{d}}^{(j)}\nonumber\\
&\cdot J\left(\sqrt{(d_j-1)J^{-1}(1-I_{v,l}^{C_j})^2+\sum_{i\ne
j}d_iJ^{-1}(1-I_{v,l}^{C_i})^2}\right).\label{Ic:irreg}
\end{align}

Combining Eqs. (\ref{Iv:irreg}) and (\ref{Ic:irreg}), one can
summarize the density evolution as a function of the mutual
information of the previous iteration, the mutual information
contribution from the other classes, noise variance, and degree
distributions
\begin{equation}
I_l^{C_j}=F(\lambda^{(j)}(x),\rho_\textbf{d}^{(j)},\sigma^2,I_{l-1}^{C_j},I_{l-1}^{C_i}),
\label{DE}
\end{equation}
where $i\in(1,\ldots,m_e)$ and $i\neq j$. By means of Eq.
(\ref{DE}), we can predict the convergence behavior of the decoding
and than optimize the degree distribution $\rho_\textbf{d}^{(j)}$
under the constraint that the mutual information shall be increasing
as the number of iterations grow, i.e.,
\begin{equation}
F(\lambda^{(j)}(x),\rho_\textbf{d}^{(j)},\sigma^2,I^{C_j},I^{C_i})>I^{C_j}.
\end{equation}

At this point, it is worth noting that, as pointed out in
\cite{plotkin:UEP} and \cite{ldpcuep}, the UEP capabilities of a
code depend on the amount of connection among the protection
classes, i.e., if the most protected class is well connected to the
least protected one, the performance of the former will decrease
while the performance of the latter will be improved. For example,
suppose a code with 2 protection classes and $d_{c_{max}}=4$. The
possible values for $\textbf{d}=(d_1,d_2)$ are (0,4), (1,3), (2,2),
(3,1), and (4,0). On the one hand, if a code has a majority of check
nodes with $\textbf{d}=(4,0)$, the first protection class will be
very isolated from class 2 which will lead to an enhanced
performance difference between the two classes. On the other hand,
if a large amount of the check nodes are of type $\textbf{d}=(2,2)$,
one can expect the protection classes to be very connected, which
favors the overall performance but mitigates the UEP capability of a
code at a moderate to large number of decoding iterations. This
indicates that for controlling the UEP capability of an LDPC code
and to prevent this characteristic from vanishing as the number of
decoding iterations grows, one has to control the amount of check
nodes of each type, i.e., optimize $\rho_\textbf{d}^{(j)}$.

\subsection{Optimization Algorithm}
The algorithm described here aims at optimizing the connection
profile between the various protection classes present on a UEP-LDPC
code, i.e., $\rho_\textbf{d}^{(j)}$. Initially, the algorithm
computes the variable node degree distribution of each class
$\lambda^{(j)}(x)$ based on $\lambda(x)$, $d_{c_{max}}$, and the
number of edges on each class. The algorithm then proceeds
sequentially optimizing the connection profile to the check nodes of
one class at a time, proceeding from the less protected class to the
most protected one. This scheduling is done in order to control the
amount of messages coming up from the less protected classes that
are forwarded to the more protected ones.

Since we are using linear programming (LP) with only a single
objective function, we chose it to be the minimization of the
average check node degree within the class being optimized, i.e., it
minimizes the average number of edges of such a class connected to
the check nodes. This minimization aims at diminishing the amount of
unreliable messages (i.e., the ones coming up from the less
protected variable nodes) that flows through a check node. In
addition to it, we control the proportion of check nodes of type
$\mathbf{d}$ introducing the parameter
$max{\rho_{\textbf{d}}^{(j)}}$ which is an upper bound on
$\sum_{\textbf{d}:d_j=s}\rho_{\textbf{d}}^{(j)}$, i.e., it limits
the proportion of check nodes of type $\textbf{d}:d_j=s$ for
$s=0,\ldots, d_{c_{max}}$, thus regulating the degree of connection
among the protection classes. For example, setting
$max{\rho_{\textbf{d}}^{(j)}}=0.35$ means that the maximum fraction
of check nodes with edge degree vector $\textbf{d}$ (for any
$\textbf{d}$) will be 0.35.

The optimization is then performed for each class $C_j$ by
minimizing its average check node degree for a decreasing
$d_{min}^{(j)}$ from $d_{c_{max}}$ to 1, where $d_{min}^{(j)}$ is
the minimum number of edges of class $j$ connected to a check node.
At this point, one can argue that since our goal is to minimize the
average connection degree within a protection class, we should thus
set $d_{min}^{(j)}=1$. The problem with this strategy is that it
would shorten the degree of freedom for the optimization of the next
class, e.g., suppose the optimization of a two-class code with
$d_{min}^{(2)}=3$ and $d_{c_{max}}=5$. Once we proceed to the
optimization of class two, the coefficients $\rho_{(2,3)}^{(2)}$,
$\rho_{(1,4)}^{(2)}$´, and $\rho_{(0,5)}^{(2)}$ are determined and
consequently fixed for the next optimization step, i.e., the
optimization of class 1, we will have as variables only the
coefficients $\rho_{(3,2)}^{(1)}$, $\rho_{(4,1)}^{(1)}$, and
$\rho_{(5,0)}^{(1)}$. Note that in this case, if we had set
$d_{min}^{(2)}=1$, there will be no degree of freedom for optimizing
class 1 since the only non-optimized $\mathbf{d}$ would be
$\mathbf{d}=(5,0)$ which would be determined by
$\sum_{s=d_{min}^{(j)}}^{d_{c_{max}}}\sum_{\textbf{d}:d_j=s}\rho_{\textbf{d}}^{(j)}=1$,
i.e., the sum of all fraction of edges must be equal to one.

\begin{figure}[!h]
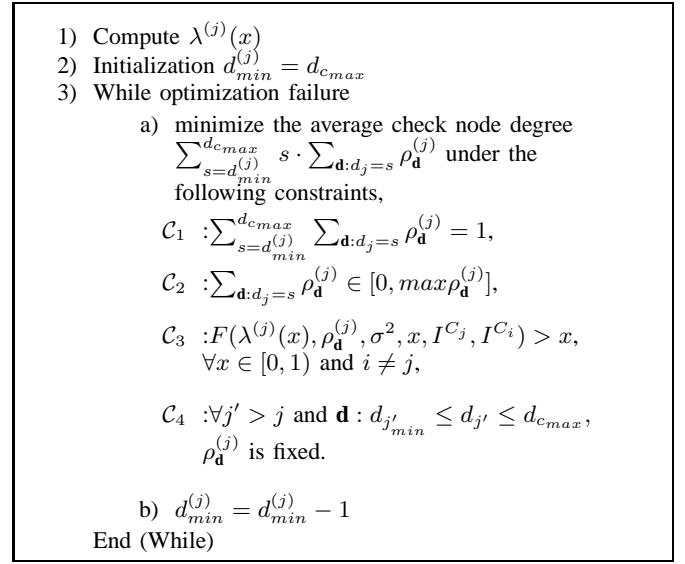

\framebox{
\begin{minipage}[b]{8.3cm}\hfill
\begin{small}
\begin{enumerate}
\item Compute $\lambda^{(j)}(x)$
\item Initialization $d_{min}^{(j)}=d_{c_{max}}$
\item While optimization failure
    \begin{enumerate}
        \item
        \begin{itemize}minimize the average check node degree\\
        $\sum_{s=d_{min}^{(j)}}^{d_{c_{max}}}s\cdot\sum_{\textbf{d}:d_j=s}\rho_{\textbf{d}}^{(j)}$
        under the\\ following constraints,
            \item[]\begin{itemize}
            \item[$\mathcal{C}_1$]:$\sum_{s=d_{min}^{(j)}}^{d_{c_{max}}}\sum_{\textbf{d}:d_j=s}\rho_{\textbf{d}}^{(j)}=1$,
            \item[]
            \item[$\mathcal{C}_2$]:$\sum_{\textbf{d}:d_j=s}\rho_{\textbf{d}}^{(j)}
                \in [0,max\rho_{\textbf{d}}^{(j)}]$,
            \item[]
            \item[$\mathcal{C}_3$]:$F(\lambda^{(j)}(x),\rho_{\textbf{d}}^{(j)},\sigma^2,x,I^{C_j},I^{C_i})>x$,
            \item[]$\forall x \in [0,1)$ and $i\neq j$,
            \item[]
            \item[$\mathcal{C}_4$]:$\forall j'>j$ and $\textbf{d}:
                d_{j'_{min}}\le d_{j'} \le d_{c_{max}},\\ \rho_{\textbf{d}}^{(j)}$ is fixed.
            \item[]
            \end{itemize}
        \item[b)]$d_{min}^{(j)}=d_{min}^{(j)}-1$
        \end{itemize}
    \end{enumerate}
\item[] End (While)
\end{enumerate}
\end{small}
\end{minipage}
}\caption{Check node profile optimization algorithm.}
\label{algorithm}
\end{figure}

The iterative procedure is successful, when a solution
$\rho_{\textbf{d}}^{(j)}$ is found which converges for the given
$\sigma^2$ and $d_{min}^{(j)}>0$. We assume that the optimizations
for classes $\{C_{j'}, j' < j\}$ have already been performed and the
results of these optimizations are used as constraints in the
current optimization process. The optimization algorithm can be
written, for given $\lambda(x)$, $\sigma^2$, $m_e$, $d_{c_{max}}$,
and $max\rho_{\textbf{d}}^{(j)}$ for $j=1,\ldots,m_e$ as shown in
Fig. \ref{algorithm}.

Note that the optimization can be solved by linear programming since
the cost function and the constraints ($\mathcal{C}_1$),
($\mathcal{C}_2$), and ($\mathcal{C}_3$) are linear in the
parameters $\rho_{\textbf{d}}^{(j)}$. The constraint
($\mathcal{C}_4$) is the previous optimization constraint. Once we
have optimized the check node profile, the code can be realized
through the construction of a parity check matrix following the
desired profile.

\section{Simulation Results}
In this section, simulation results for multi-edge type UEP-LDPC
codes with optimized check node connection profile are presented. We
designed UEP-LDPC codes of length $n=4096$ with $m_e=3$ protection
classes, rate 1/2, and $d_{v_{max}}=30$ following the algorithm of
\cite{Poulliat:UEPLDPC}. The proportions of the classes are chosen
such that $C_1$ contains 20\% of the information bits and $C_2$
contains 80\%. The third protection class $C_3$ contains all parity
bits. Therefore, we are mainly interested in the performances of
classes $C_1$ and $C_2$. The optimized variable and check node
degree distribution for the UEP-LDPC code are given by
$\lambda(x)=0.2130x+0.0927x^2+0.2511x^3+0.2521x^{17}+0.0965x^{18}+0.0946x^{29}$
and $\rho(x)=x^8$, respectively.

In order to have a low-complexity systematic encoder, we construct
parity check matrices in lower triangular form
\cite{richardson:efficientenc}. This approach also leads to a
simplification in our optimization procedure, i.e., given that the
parity bits are in the less protected class $C_3$ and that they
should be organized in a lower triangular form, we start the
optimization from the less protected information bits class $C_2$,
since all the connections between the variable nodes of class $C_3$
and the check nodes are completely determined by the lower
triangular form construction algorithm. Table \ref{lambda:local}
summarizes the classes' variable degree distributions
$\lambda^{(j)}(x)$.
\begin{table}
\centering \caption{Local variable degree distributions. The
coefficients $\lambda_i^{(j)}$ represent the fraction of edges
connected to variables nodes of degree $i$ within the class $C_j$.}
\scalebox{0.97}{
    \begin{tabular}{|c|c|c|}
    \hline
    $C_1$ & $C_2$ & $C_3$ \\\hline
    $\lambda_4^{(1)}=0.00197$  & $\lambda_3^{(2)}=0.23982$ & $\lambda_2^{(3)}=0.93901$ \\
    $\lambda_{18}^{(1)}=0.57263$ & $\lambda_4^{(2)}=0.76018$ & $\lambda_3^{(3)}=0.06099$ \\
    $\lambda_{19}^{(1)}=0.21085$ & ~                & ~ \\
    $\lambda_{30}^{(1)}=0.21455$ & ~                & ~ \\
    \hline
    \end{tabular}}
\label{lambda:local}
\end{table}
We applied the optimization algorithm for different values of
$max\rho_{\textbf{d}}^{(j)}$ to enable the observation of the
varying UEP capabilities of the codes. The resulting distributions
are summarized in Table \ref{R:opt}. All the simulations were done
for a total of 50 decoding iterations and the constructed codes were
all realized through a modification of progressive edge-growth (PEG)
\cite{PEG} algorithm done in order to ensure that the optimized
check node degree is realized.
\begin{table}
\centering \caption{Optimized check node profile for 3 protection
classes. The coefficients $\rho_s^{(j)}$ represent
$\rho_{\textbf{d}}^{(j)}$ with $\textbf{d}:d_j=s$.} \scalebox{0.97}{
    \begin{tabular}{|c|c|c|c|}
    \hline
    ~&$C_1$ & $C_2$ & $C_3$\\\hline
    $max\rho_{\textbf{d}}^{(2)}=0.35$&$\rho_2^{(1)}=0.08977$ &$\rho_2^{(2)}=0.35$&$\rho_1^{(3)}=0.00024$\\
    ~                               &$\rho_3^{(1)}=0.19637$  &$\rho_3^{(2)}=0.35$&$\rho_2^{(3)}=0.93877$\\
    ~                               &$\rho_4^{(1)}=0.34911$  &$\rho_4^{(2)}=0.30$ &$\rho_3^{(3)}=0.06099$\\
    ~                               &$\rho_7^{(1)}=0.36475$  &      ~            & ~
    \\
    \hline
    $max\rho_{\textbf{d}}^{(2)}=0.55$&$\rho_3^{(1)}=0.25248$&$\rho_3^{(2)}=0.45$ &$\rho_1^{(3)}=0.00024 $\\
    ~                               &$\rho_4^{(1)}=0.54860$  &$\rho_4^{(2)}=0.55$&$\rho_2^{(3)}=0.93877$\\
    ~                               &$\rho_7^{(1)}=0.19892$  & ~                 &$\rho_3^{(3)}=0.06099$\\
    \hline
    $max\rho_{\textbf{d}}^{(2)}=0.75$&$\rho_3^{(1)}=0.14027$&$\rho_3^{(2)}=0.25$  &$\rho_1^{(3)}=0.00024 $\\
    ~                               &$\rho_4^{(1)}=0.74809$&$\rho_4^{(2)}=0.75$  &$\rho_2^{(3)}=0.93877$\\
    ~                               &$\rho_7^{(1)}=0.11164$&          ~           &$\rho_3^{(3)}=0.06099$\\
    \hline
    \end{tabular}}
\label{R:opt}
\end{table}

\begin{figure}[h]
  \centering
  \hspace*{-3.9mm}
  \includegraphics[scale=.69]{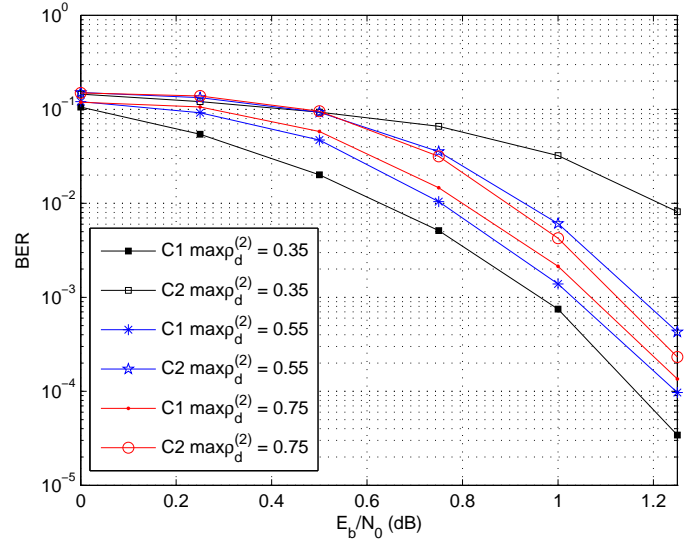}
  \caption{Classes bit error rate of the optimized multi-edge unequal error protection LDPC codes for different values of
  $max\rho_{\textbf{d}}^{(2)}$.}\label{results}
\end{figure}

Figure \ref{results} shows that the difference in the performances
of the protection classes is reduced as we increase the value of
$max\rho_{\textbf{d}}^{(2)}$. This is an expected effect, since the
greater $max\rho_{\textbf{d}}^{(2)}$, the greater is the amount of
information that $C_2$ exchanges with $C_1$. Obviously, this is
expected to enhance the performance of $C_2$ while lowering the one
of $C_1$. Furthermore, despite the already large number of decoding
iterations (50 iterations), the UEP capability is preserved,
something regarded as infeasible in \cite{Poulliat:UEPLDPC} and not
observed in \cite{ldpcuep} for codes realized by means of PEG.
\section{Concluding remarks}
In this paper, we introduced a multi-edge type analysis of unequal
error protection LDPC codes. By means of such an analysis, we
derived an optimization algorithm that aims at optimizing the
connection profile between the protection classes within a codeword.
This optimization allowed us not only to control the differences in
the performances of the protection classes by means of a single
parameter, but also to prevent the UEP capability of an LDPC code to
vanish after a moderate to large number of decoding iterations.

\section*{Acknowledgment}
This work is funded by the German Research Foundation (DFG).

\bibliographystyle{IEEEtran}
{\small
 \bibliography{ref}
}

\end{document}